\documentstyle[preprint,aps,prb,epsf]{revtex}

\begin{document}
\draft
\title{Surface Magnetic Phase Diagram of Tetragonal Manganites}

\author{Zhong Fang$^{1}$, and Kiyoyuki Terakura$^{2,3}$}
\address{
$^{1}$JRCAT, Angstrom Technology Partnership, 1-1-4 Higashi, Tsukuba,
Ibaraki 305-0046, Japan \\
$^{2}$JRCAT, National Institute for Advanced Interdisciplinary Research,
        1-1-4 Higashi, Tsukuba, Ibaraki 305-8562, Japan \\
$^{3}$Tsukuba Advanced Computing Center, 1-1-4 Higashi, Tsukuba,
Ibaraki 305-8561, Japan
}

\maketitle

\begin{abstract}
  To gain insights into the fundamental and characteristic features of
  the surface of doped manganites, we constructed a general magnetic
  phase diagram of La$_{1-x}$Sr$_{x}$MnO$_3$ (001) surfaces in the
  plane spanned by $x$ and the bulk tetragonal distortion $c/a$, from
  the first-principles calculations. We found that the surfaces are
  quite different from the bulk in the sense that both the (La, Sr)O
  and MnO$_2$ terminated surfaces show strong tendency toward
  antiferromagnetism (A-type and C-type respectively).  The basic
  physics governing the phase diagram can be understood in terms of
  the surface orbital polarizations.  It is also found that the strong
  surface segregation of Sr atoms is mostly caused by the
  electrostatic interaction and will further enhance the tendency to
  surface antiferromagnetism.
\end{abstract}

\newpage

\section{Introduction}

Over the last half a decade, the perovskite colossal magnetoresistive
(CMR) manganites have attracted intensive attention due to the
remarkably rich variety of structural, magnetic, transport and optical
properties~\cite{CMR} and are regarded as potentially important
materials for the next generation technology.  Such possibility may be
a strong motivation for the study of thin films and
superlattices~\cite{Izumi} of the manganites and the related
materials.  In these systems surfaces and interfaces will play
important roles.  However, even without appealing to thin films and
superlattices, surfaces and interfaces are involved in various
aspects.  The enhanced low-field magnetoresistance has been reported
in the polycrystalline samples due to the spin-dependent behaviors
across the grain boundaries~\cite{IMR}. The photoemission
data~\cite{PES}, which provide evidence for the half-metallicity, and
the scanning tunneling spectroscopy (STS)~\cite{STS} or microscopy
(STM)~\cite{STM}, which provide evidence for the spatial
phase-separation~\cite{phase-separation}, are all surface
sensitive. Therefore, the basic understanding of the surfaces of doped
manganites is an urgent and challenging problem. It is already
established that one of the most important implications in the physics
associated with the CMR manganites is the key roles of orbital degrees
of freedom (ODF)~\cite{ODF}, which couple strongly with the lattice,
charge and magnetic degrees of freedom (MDF).  We expect that the ODF
will play even more important roles on the surfaces due to the
lowering in symmetry and dimensionality.

Detailed experimental and theoretical studies on the surfaces of CMR
manganites have started only very recently~\cite{Ma,Peng,Pickett,Brey,LCMO}
and the results obtained so far are not yet sufficient to establish
any general picture about the surface phase diagram.  In the
experimental side, it is still difficult to prepare well defined
surfaces. In the theoretical side, on the other hand, only some
limited phase space and limited conditions have been taken into
account.  Although it is not possible to consider the complete phase
space and all possible complications conceivable in real systems, it
is very important to construct a general qualitative picture for the
surface phase diagram by taking account of possible phases and some
important experimental conditions.  This is what we aim in the present
work by taking La$_{1-x}$Sr$_x$MnO$_3$ (LSMO) as a canonical system of
doped manganites~\cite{CE}.

First, since it is possible to control the terminations by chemical
treatment of the surface~\cite{kawasaki} or by controlling the
chemical potential~\cite{BTO}, consideration of two possible
terminations of (001) surface is important. It is natural to expect
that the surface effect will be strong for the MnO$_2$ termination,
but weak for the (La, Sr)O termination, due to the loss of part of the
ligand of Mn in former case.  Although the expectation is
qualitatively true, the change in the electrostatic potential for SrO
termination can also produce significant change in the electronic
structure. Second, the lattice deformation induced by the substrates
can produce cooperative stabilization of orderings in ODF and MDF,
leading to strong anisotropy in the thin film~\cite{Konishi,Phase}.
Therefore, in the present work, we will present a general surface
phase diagram of LSMO, for both the (La, Sr)O and MnO$_2$ terminated
(001) surfaces, as functions of the hole doping $x$ and the bulk
tetragonal distortion $c/a$ ratio induced by the substrates.  We found
that the (La, Sr)O terminated surface shows strong tendency toward the
A-type antiferromagnetic (AF) state, while the surface phases of
MnO$_2$ termination are dominated by the C-type AF state. The basic
physics behind is the ordering in the surface ODF of Mn $e_g$ states.
Third, we found that the Sr atoms should seriously segregate towards
the surfaces due to the electrostatic interaction. The general
tendency of surface segregation will further favor the surface AF
states.

\section{Calculation Methods}

The present work is based on the first-principles electronic structure
calculations, which adopt the PBE version~\cite{PBE} of generalized
gradient approximation (GGA). The Kohn-Sham equation is solved by
using the pseudopotential technique. The Mn(3d) and the O(2p) states
are treated by the ultra-soft pseudopotential~\cite{USPP}, while the
norm-conserving scheme~\cite{NCPP} is used for other states.  The
cutoff energies for the plane-wave expansions of wavefunction and
charge density are 30 Ry and 200 Ry respectively. La$_x$Sr$_{1-x}$ is
treated as a virtual atom by the virtual crystal approximation
(VCA)~\cite{VCA}. In present systems, the valence electronic states
near and below the Fermi level are contributed mostly by Mn and O, the
states coming from La and Sr are far above the Fermi level. Therefore,
the simple VCA can provide a reasonable description of the
systems. The validity of these techniques was well demonstrated in our
previous calculations for the bulk~\cite{Phase}. We use a repeated
slab geometry, which includes five MnO$_2$ layers with mirror symmetry
in the central MnO$_2$ layer.  As the MnO$_2$ layer and the (La, Sr)O
layer are stacked alternately along the surface normal, i.e.,
$z$-axis, nine and eleven atomic layers in total are included in the
unit cell for the MnO$_2$ and (La, Sr)O terminations, respectively.  A
vacuum region of $12\sim 13$\AA\ is included to separate the slabs.
Several magnetic states, i.e., ferromagnetic (FM), A-type AF and
C-type AF states, are calculated in the present study. In the A-type
AF state, the magnetic moment of Mn are aligned ferromagnetically in
the $ab$-plane (slab plane) and these FM layers are coupled
antiferromagnetically along the $c$ direction (surface normal). In the
C-type AF state, on the other hand, the FM chains along the $c$
direction are coupled antiferromagnetically.  In the surface phase
diagrams in Fig.1, the above definition is applied to the slab. A
simple schematic description about the spin configurations of surface
A- and C-type AF states can be found in Fig.2.  To accommodate the
C-type AF states, we use the planar c(2x2) unit cell. The atomic
positions and magnetic states are fully optimized for each of given
$ab$-plane lattice constants, which define the bulk $c/a$ ratio since
the bulk volume is given for each doping $x$.  By choosing the lowest
energy state among FM, A-type AF and C-type AF states for each
$ab$-plane lattice constant and each doping $x$, we can construct the
surface phase diagram.

\section{Results and discussion}

\begin{figure}[t]
      \centering
      \leavevmode  \epsfxsize=120mm \epsfbox[120 230 450 530]{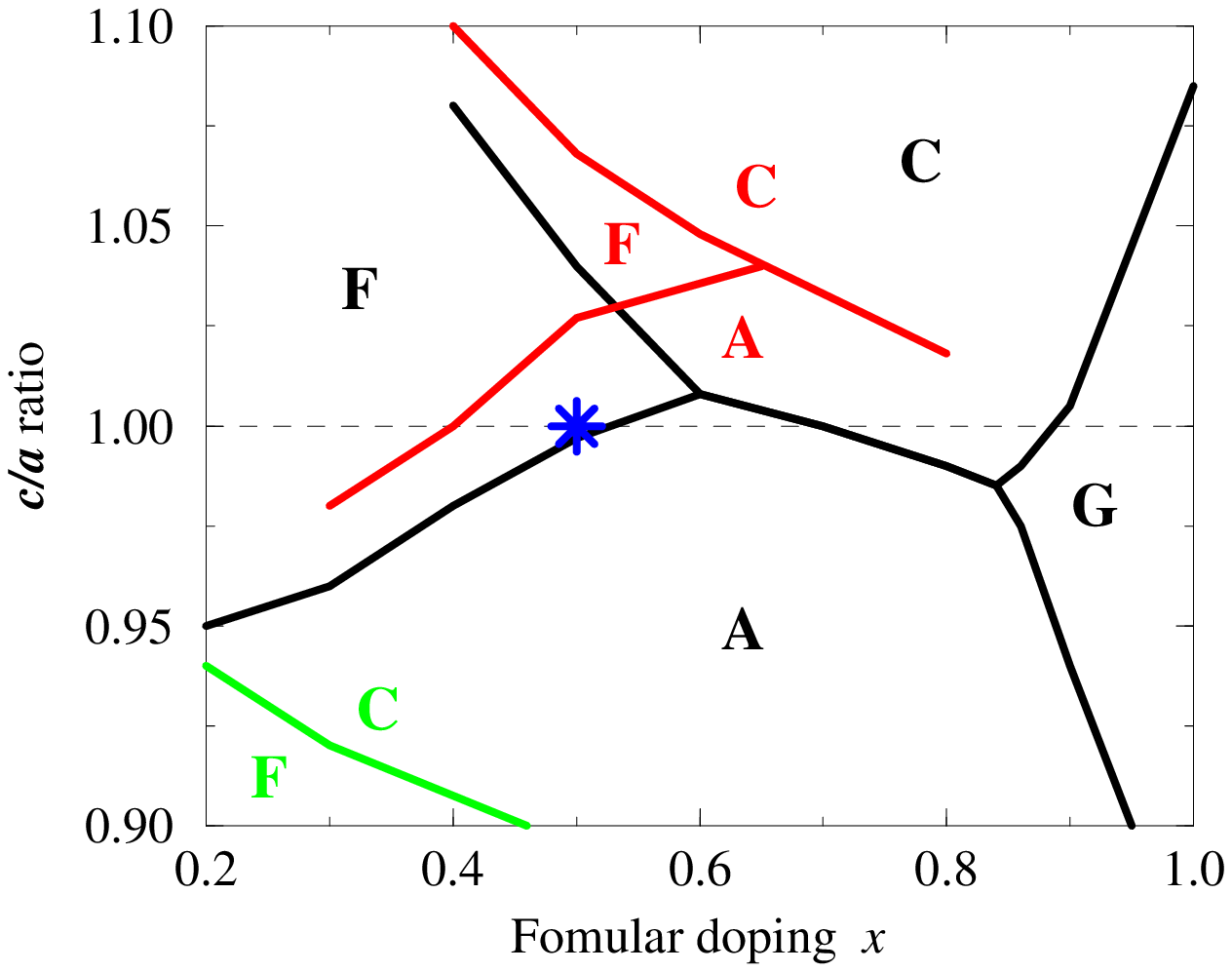}
      \caption{The calculated phase diagrams of La$_{1-x}$Sr$_x$MnO$_3$
        in a plane of hole doping $x$ and $c/a$ ratio. Here the $c/a$
        ratio can be defined by the $ab$-plane lattice constant as the
        bulk volume is given for each doping $x$. The red and green
        colors correspond to the surface phase diagrams for the (La,
        Sr)O and the MnO$_2$ terminations respectively, while the
        black color corresponds to the bulk phase diagram. The
        denotations F, A , C and G mean FM, A-type AF, C-type AF and
        G-type AF states respectively.}
      \label{figure:phase}
\end{figure}

We show the calculated surface phase diagrams in
Fig.~\ref{figure:phase} compared with the bulk phase diagram which was
obtained in our previous calculations~\cite{Phase}.  The surface phase
diagrams for the (La, Sr)O and MnO$_2$ terminations are indicated by
the red and green lines respectively.  Clearly for the MnO$_2$
termination, the surface phases are dominated by the C-type AF state
for a wide range of doping $x$ and $c/a$ ratio, except a tiny FM
region in the very small $c/a$ ratio and low doping case.  On the
other hand for the (La, Sr)O termination, the surface shows rich
phases similar to the bulk case but with a significant shift.
Compared with the bulk phase diagram (indicated by black lines and
characters), the main change for the surfaces is the up- and downward
shifts of the phase boundaries for (La, Sr)O and MnO$_2$ termination
respectively.  It should be noted that, since non-equal numbers of
(La, Sr)O and MnO$_2$ layers are included in our unit cell, the
effective doping $x_{\rm eff}$ of the central MnO$_2$ layer is
slightly different from the formula doping $x$.  By calculating the
number of $e_g$ electrons, we estimate that the $x_{\rm eff}$ is
smaller (larger) than $x$ by about 0.04 (0.05) for (La, Sr)O (MnO$_2$)
termination.  For the sampling point ($x=0.5$ and $c/a=1.0$, indicated
by the blue star in Fig.~\ref{figure:phase}), the calculated total
energies and some physical parameters are summarized in Tables I and
II. It should be also noted here that the presence of interfaces
between the surface and bulk regions may modify our calculated phase
diagram slightly. If the penetration of the perturbation in the
magnetic state on surface is deeper than half of the slab thickness,
the surface phase diagram may depend on the slab thickness. In order
to check these effects, we performed calculations in which the spin
structures of the central MnO$_2$ layer are constrained to be the bulk
ones. We found that the spin structures in the first two surface
layers (of both sides of the slabs) are little affected by the
constraint in the central layer. We should also point out that the
possible non-collinear magnetic configurations are not taken into
account in the present calculation. However, even if we confine
ourselves to collinear spin configurations, we can still gain
important valuable insights into the basic trend in surface magnetic
states by our calculated phase diagram.

The occupation imbalance between two $e_g$ orbitals, namely $3z^2-r^2$
and $x^2-y^2$ orbitals in doped manganites, is the key concept to
understand the rich phase diagrams~\cite{CMR,Phase,Igor}. On the
surfaces, the orbital polarization is certainly very important because
of the change in crystal field, hybridization and surface lattice
relaxation. For the (La, Sr)O termination, the environment of the 2nd
surface Mn changes only from the 2nd nearest neighbor layer, which is
negatively charged MnO$_2$ layer.  Missing of this layer in the vacuum
side produces an attractive potential leading to downward shift of the
2nd surface Mn $3d$ bands. However, the strengthened $pd\sigma$
hybridization due to the missing Mn atom on top of the surface O atom
will cause the upward shift of the $3z^2-r^2$ state compared with the
$x^2-y^2$ state.  Although the surface relaxation in the interlayer
distance, which is very small, will modify the details quantitatively,
it will not affect the qualitative features. For the sampling
structure in Fig.~1, the Mn-O bond length between the 1st and 2nd
layers changes only by $-0.2\%$ (see Table II). For the MnO$_2$
termination, the most dramatic effect is the significant downward
shift of the $3z^2-r^2$ state due to the reduction in $pd\sigma$
hybridization caused by the missing ontop oxygen.  For other orbitals,
missing of the positively charged (La, Sr)O layer produces repulsive
potential.  In this case, the surface relaxation is very large because
of the loss of ligand oxygen atoms.  However, the atoms are relaxed in
such a way that the topmost Mn-O bond length along surface normal
direction is elongated to further push down the $3z^2-r^2$ orbital.
Therefore the net effect of surface orbital polarization is that the
occupation of the $x^2-y^2$ ($3z^2-r^2$) orbital is enhanced for (La,
Sr)O (MnO$_2$) termination.  The orbital population ratio $n_{\rm
  3z^2-r^2}/n_{\rm x^2-y^2}$ for the sampling structures in Fig.~1 are
estimated as 0.74 and 2.16 for the surface Mn sites of (La, Sr)O and
MnO$_2$ terminations respectively. The surface orbital polarization of
MnO$_2$ termination is much stronger than that of (La, Sr)O
termination.  Figure 2 shows the charge distributions (from the Fermi
level to 0.8 eV below) for two different terminations of the sampling
point. The surface orbital polarizations are clearly seen in each
case.

The change in occupation will change the competition between the
double exchange (DE) and the superexchange (SE). For the less than
half-filled majority-spin $e_g$ bands, the more (less) populated are
the orbitals, the stronger (weaker) are the DE interactions among
these orbitals.  Therefore, the region of the A-type (C-type) AF state
becomes wider for the (La, Sr)O (MnO$_2$) termination because of
higher population of the $x^2-y^2$ ($3z^2-r^2$) orbital.
A. Filippetti and W. E. Pickett~\cite{Pickett} studied the magnetic
properties of MnO$_2$ terminated (001) surface of
La$_{1-x}$Ca$_x$MnO$_3$ (LCMO) with particular doping $x$
(=0.5). Their observation of the stability of the FM state is due to
the use of the (1x1) surface unit cell.  Our calculations for LSMO
with the same doping ($x=0.5$, $c/a=1.0$) give the same order of total
energy difference between the surface FM and A-type AF states (Table
I). However, the surface C-type AF state has much lower energy
compared with the above two states.

On the real surfaces of CMR manganites, the surface segregation on the
perovskite A-sites may be another important factor. It was
experimentally suggested that the Ca content in the surface layers of
LCMO is dramatically enhanced~\cite{LCMO} for both terminations. The
basic questions here are: 1)why does surface segregation happen? 2)how
will it affect our phase diagram?  In the analysis of surface
segregation, we need the bulk part as a particle reservoir.  The
present type slab calculation has, therefore, rather severe
restriction in this context.  In the following, we will do a simple
analysis in order to gain insight into the fundamental aspects of the
problem.  We artificially locate a pure SrO layer at different
positions in our unit cell keeping the Sr content $x$ in all other
(La, Sr)O layers as a given value and calculate the total energy as a
function of the SrO layer position. The calculated results shown in
Fig. 3 clearly suggest that the stable position of the SrO layer is
the surface for the (La, Sr)O termination or the 2nd surface for the
MnO$_2$ termination.  The result is qualitatively consistent with the
experimental observation for LCMO~\cite{LCMO}.  The main reason for
the stability of the surface (or subsurface) SrO layer is the
electrostatic interaction.  SrO layer is nominally charge neutral,
while the nominal charge of LaO is +1.  In the bulk the ionized object
is stabilized by the electrostatic interaction with the counter ions.
This stability mechanism is weakened on the surface.  Therefore, the
neutral object tends to be located on the surface keeping the charged
objects inside the bulk.  For the MnO$_2$ termination, the presence of
SrO layer at the 2nd surface will make the surface MnO$_2$ layer
nearly neutral also.  In Fig.~3(b), we show that the contribution
coming from the electrostatic energy is responsible to the stability
of the SrO layer at the surface.  As the nominal doping $x$ increases,
the effect of Sr segregation should be weakened as is actually
demonstrated in Fig.~3.  Although the present analysis is only for
limited configurations in which one particular layer of the system has
full Sr segregation.  Nevertheless, the above consideration about the
mechanism of the stability of surface SrO layer clearly suggests a
rather general tendency of surface segregation of the neutral objects.
By taking account of the surface segregation of Sr, the effective
doping on the surface becomes larger than the bulk doping. This will
again lead to stronger stability of surface AF ordering.

\section{Summary}

In summary, we calculated the surface magnetic phase diagram of
tetragonal manganite LSMO as functions of hole doping $x$ and the bulk
tetragonal distortion $c/a$.  The (001) surfaces of tetragonal
manganites show clear tendency towards A-type and C-type AF states for
(La, Sr)O and MnO$_2$ terminations respectively though the surface
A-type AF state may be hard to distinguish experimentally from the
surface FM state.  The basic physics governing the phase diagram is
explained in terms of the orbital polarization induced by surface
effects. Strong surface segregation of Sr atoms is caused by the
electrostatic interaction and further favors the tendency to surface
antiferromagnetism.

\section{Acknowledgments}

The authors thank Professors Y. Tokura, E. W. Plummer, M. Kawasaki and
R. Matzdorf for valuable comments and for providing us with their
experimental data. The present work is partly supported by NEDO.

\newpage

\newpage
\begin{figure}[t]
      \centering
      \leavevmode  \epsfxsize=120mm \epsfbox[90 20 880 680]{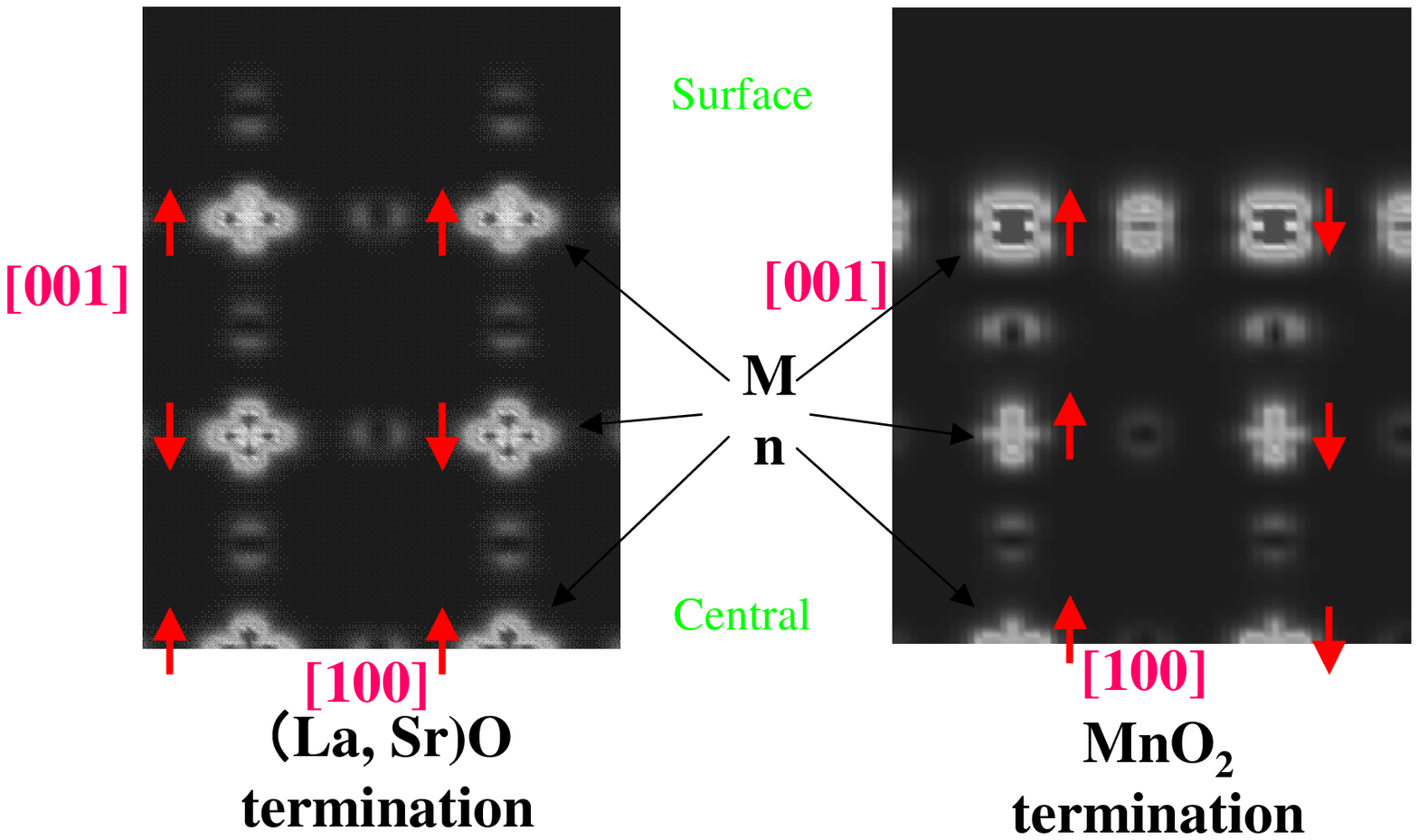}
      \caption{The occupied (0.8 eV below the Fermi level) electronic
        charge distributions for the two terminations corresponding to
        the sampling point in Fig. 1. Red color denotes high charge
        density.  Between two neighboring Mn sites, there is small
        charge density from oxygen sites. The spin configurations of
        Mn sites are indicated by the red arrows.}
      \label{figure:charge}
\end{figure}

\newpage
\begin{figure}[t]
      \centering
      \leavevmode  \epsfxsize=120mm \epsfbox[100 100 500 600]{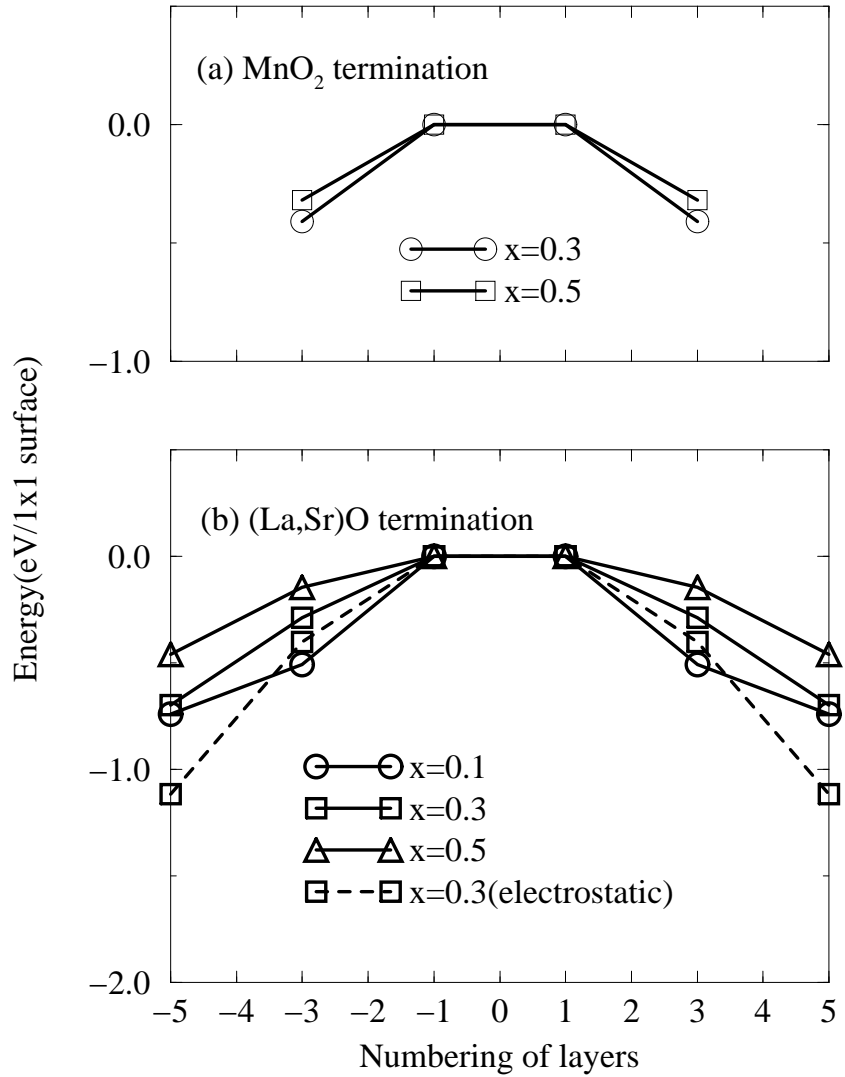}
      \caption{The calculated total energies as a function of SrO layer
        position keeping the Sr content $x$ in all other (La, Sr)O
        layers as a given value. The zero in the numbering of layers
        corresponds to the center of our unit cell, while MnO$_2$
        layers are located on even number of layers.}
\end{figure}

\begin{table}[t]
\caption{The calculated total energies, per surface (1x1) cell,
    for the sampling point ($x=0.5$, $c/a=1.0$) in Fig.~1.}
\begin{tabular}{c|c|c|c}
    &Surface F   &Surface A  &Surface C  \\  \hline
(La,Sr)O termination  &22 meV  &0 meV   &61 meV    \\
MnO$_2$ termination &124 meV  &203meV  &0 meV  \\
\end{tabular}
\end{table}

\begin{table}[t]
\caption{Some calculated parameters corresponding to the sampling point
  in Fig.~1 for the two terminations.  $\Delta d_{\rm Mn-O}$ means the
  change of the topmost Mn-O bond length along surface normal; $M_{\rm c}$
  and $M_{\rm s}$ denote the magnetic moments of Mn sites at the central and
  the surface layers respectively; $n_{\rm 3z^2-r^2}/n_{\rm x^2-y^2}$
  defines the ratio of occupation numbers for two $e_g$ orbitals of
  surface Mn.}
\begin{tabular}{c|c|c|c|c}
    &$\Delta d_{\rm Mn-O}$ &$M_{\rm c}$ &$M_{\rm s}$ &$n_{\rm 3z^2-r^2}/n_{\rm x^2-y^2}$ \\
    \hline (La,Sr)O A-AF &-0.2\% &3.11 &3.12 &0.74 \\ MnO$_2$ C-AF
    &+2.7\% &2.87 &3.17 &2.16 \\
\end{tabular}
\end{table}


\begin{thebibliography}{50}
\bibitem{CE} Since the tilt of MnO$_6$ octahedra in
    La$_{1-x}$Sr$_x$MnO$_3$ is rather small, we can treat as a first
    approximation the system as tetragonal. Therefore the complicated
    charge ordering state does not occur. It is demonstrated in our
    previous study for the bulk that the essential physics with respect
    to the ODF can be caught in this simple tetragonal lattice.
\end{thebibliography}
\end{document}